\documentclass[conference]{IEEEtran}

\usepackage{amsfonts}
\usepackage{amssymb}
\usepackage{subfigure}
\usepackage{amsmath}
\usepackage{amsthm}
\usepackage{mathrsfs}
\usepackage{verbatim}
\usepackage{graphicx}
\usepackage{epstopdf}
\usepackage{enumitem}
\usepackage{caption}
\usepackage{xcolor}
\usepackage{cite}

\theoremstyle{remark} \newtheorem{remark}{Remark}

\newtheorem{lemma}{Lemma}
\newtheorem{proposition}{Proposition}
\newtheorem{definition}{Definition}

\newtheorem{corollary}{Corollary}

\theoremstyle{remark} \newtheorem{theorem}{Theorem}

\newcommand{\yvec}{\mathbf{y}}

\newcommand{\mX}{\mathcal{X}}

\newcommand{\realSet}{\mathcal{R}}

\newcommand{\E}{\mathbb{E}}

\renewcommand{\vec}[1]{\mathbf{#1}}
\renewcommand{\sf}[1]{\mathsf{#1}}

\newcommand{\argmax}{\operatornamewithlimits{argmax}}

\newcommand{\changeColor}{black}
\newcommand{\ymc}[1]{{\color{black}{#1}}}

\begin{document}
\IEEEoverridecommandlockouts
\title{\ymc{Diversity Gain of One-shot Communication \\ over Molecular Timing Channels}
\thanks{This research was supported in part by the NSF Center for Science of Information (CSoI) under grant CCF-0939370, and the NSERC Postdoctoral Fellowship fund PDF-471342-2015.}
}

\author{
Yonathan Murin, Mainak Chowdhury, Nariman Farsad, and Andrea Goldsmith \\
{Department of Electrical Engineering, Stanford University, USA} \\
\vspace{-.55cm}
}
\date{}


\maketitle

\begin{abstract}
	We study \ymc{diversity in} one-shot communication over molecular timing channels. In the considered channel model the transmitter {\em simultaneously} releases a large number of information particles, where the information is encoded in the {\em time of release}. The receiver decodes the information based on the {\em random} time of arrival of the information particles. We characterize the asymptotic exponential decrease rate of the probability of error as a function of the number of released particles. \ymc{We denote this quantity as the {\em system diversity gain}, as it depends both on the number of particles transmitted as well as the receiver detection method.} Three types of detectors are considered: the maximum-likelihood (ML) detector, a linear detector, and a detector that is based on the first arrival (FA) among all the transmitted particles. 
	We show that for random propagation characterized by right-sided unimodal densities with zero mode, the FA detector is equivalent to the ML detector, and significantly outperforms the linear detector. Moreover, even for densities with positive mode, the diversity gain achieved by the FA detector is very close to that achieved by the ML detector and much higher than the gain achieved by the linear detector.
\end{abstract}

\section{Introduction}	

In many communication systems it is common to modulate the information in the amplitude or in the phase of the transmitted signal.
In this work we consider a different transmission approach in which the information is embedded in the {\em timing} of the transmissions. 
The resulting channels are commonly referred to as {\em timing channels}. 
Communication over timing channels was studied in three main contexts: communication via queues, i.e., queuing timing channels \cite{ana96, sund00-1, sund00-2, kivayash09, sellke07}, molecular communications, i.e., molecular timing channels, \cite{eck07, sri12, li14, rose:InscribedPart1, isit16, murinGlobeCom16}, and covert (secure) timing channels \cite{ulukus16, dunn09}.

We study a model for molecular timing channels where information is modulated through the {\em time of release} of information particles \textcolor{\changeColor}{(see \cite{far16ST} for a detailed discussion regarding applications of molecular communications)}. 
These information particles represent molecules in the context of molecular communications, or tokens using the terminology of \cite{rose:InscribedPart1}.
We focus on a one-shot communication scenario in which the transmitter {\em simultaneously} releases multiple identical information particles, where the time of release is selected out of a set with {\em finite} cardinality. 
The receiver's objective is to detect this time of release. The released particles are assumed to randomly and independently propagate to the receiver, where upon their arrival they are absorbed and removed from the environment. 
Thus, the random delay until a particle arrives at the receiver can be represented as an additive noise term. 
Our objective is to characterize the asymptotic exponential decrease rate of the probability of error, at the receiver, as a function of the number of released particles. 
We refer to this quantity as the {\em \ymc{system} diversity gain}.\footnote{\ymc{This quantity can also be interpreted as a function of the number of particles, in contrast to common usage of error exponent to characterize the exponential rate of decrease of the probability of error with the increase in the block length.}} \ymc{The formal definition of diversity gain is given in Section \ref{sec:ProbForm}.} 
\textcolor{\changeColor}{Note that the work \cite{asilomar16} also considered a molecular timing channel with diversity, yet that work focused on the capacity of the channel while in the current work we study the diversity gain in the probability of error for one-shot communication.}
Comparing the diversity gains of different detection techniques indicates which method achieves lower probability of error.

As we consider a causal \ymc{molecular} timing channel, we focus on propagation models characterized by noise densities with support on the positive real line. 
\ymc{In particular, in molecular communications, the particles propagate to the receiver following a random Brownian path. When the propagation is based solely on diffusion, the additive noise associated with random delay follows the L\'evy distribution \cite{murinGlobeCom16}. When the diffusion is accompanied by a drift, this additive noise follows the inverse Gaussian (IG) distribution \cite{sri12, li14}. In the model studied in \cite{rose:InscribedPart1}, the additive noise representing the propagation of the tokens follows an exponential distribution. The exponential delay can represent systems with chemical reactions that cause the particles to decay quickly \cite{guo16}.}


Motivated by the scenarios studied in \cite{sri12, li14, rose:InscribedPart1, isit16, murinGlobeCom16}, we further assume that the noise density associated with the random propagation delay is unimodal (with support on the positive real line) and derive expressions for the \ymc{system} diversity gain \ymc{associated with} three types of detectors: the optimal maximum likelihood (ML) detector, a linear detector based on the mean of the arrival times, and a detector that is based on the first arrival (FA) among the transmitted particles \cite{murinGlobeCom16}. 
One of the main results presented in \cite{murinGlobeCom16} is that in the case of a L\'evy-distributed additive noise, linear detection under multiple particle release has worse performance than linear detection based on a single particle release. 
This degradation is due to the fact that the L\'evy distribution has heavy tails that render linear processing highly suboptimal. It was further shown in \cite{murinGlobeCom16} that for a small number of released particles, the performance of the FA detector is indistinguishable from that of the ML detector; thus, \ymc{this detector provides a simple and attractive alternative to ML detection for a small number of released particles}. 

In this work we extend this result to the case of a large number of released particles. We show that if the noise density has a zero-mode, for example as is the case for uniform or exponential distributions, then the FA and ML detectors {\em are equivalent}. Moreover, even if the mode is larger than zero, the FA detector can still achieve a diversity gain {\em very close} to the one achieved by the ML detector, and can {\em significantly outperform the linear detector}. 
\ymc{This holds {\em regardless} of the tails of the noise, and contradicts the common use of linear processing, known to maximize the signal-to-noise ratio (or minimize the bit error rate) in systems with receive diversity and additive Gaussian noise \cite{TV:05}.
%
%
%
\noindent Our results indicate that for detection of signals transmitted over molecular timing channels, the FA detector is a much better alternative to the high-complexity ML detector as compared to linear processing.}

The rest of this paper is organized as follows.
The problem formulation is presented in Section \ref{sec:ProbForm}. 
The diversity gain of the ML, linear, and FA detectors is derived in Section \ref{sec:PDG}. 
Analysis of the diversity gain of specific densities, namely, the uniform, exponential, IG, and L\'evy, is provided in Section \ref{sec:numRes}, and concluding remarks are provided in Section~\ref{sec:conc}.

{\bf {\slshape Notation}:} 
We denote sets with calligraphic letters, e.g., $\mathcal{X}$, where $\realSet^{+}$ denotes the set of positive real numbers.
We denote RVs with upper case letters, e.g., $X$, and their realizations with lower case letters, e.g., $x$. 
An RV takes values in the set $\mX$, and we use $|\mX|$ to denote the cardinality of a finite set. 
We use $f_{Z}(z)$ to denote the probability density function (PDF) of a continuous RV $Z$ on $\realSet^{+}$ and $F_{Z}(z)$ to denote its cumulative distribution function (CDF). 
We denote vectors with boldface letters, e.g., $\yvec$, where the $k^{\text{th}}$ element of a vector $\yvec$ is denoted by $y_k$.
Finally, we use $\log (\cdot)$ to denote the natural logarithm.

\section{Problem Formulation} \label{sec:ProbForm}

\subsection{System Model} \label{subsec:sysModel}

We make the following assumptions about the system:\footnote{Note that these assumptions are consistent with those made in previous works \cite{eck07, sri12, li14, rose:InscribedPart1, isit16, murinGlobeCom16}.}


\begin{enumerate}[label = \emph{\roman*})]
	\item \label{assmp:timeModulation}
	The information particles are assumed to be {\em identical and indistinguishable}, thus, the information is encoded {\em only} in the time of release of the particles. At the receiver, the information is decoded based {\em only} on the time of arrival. The propagation of the particles from the transmitter to the receiver is {\em random}, inducing a random arrival time at the receiver.
	
	\item \label{assmp:synch}
	The time-synchronization between the transmitter and the receiver is perfect, the transmitter perfectly controls the particles' release time, and the receiver perfectly measures their arrival time. 
	
	\item \label{assmp:Arrival}
	Every information particle that arrives at the receiver is absorbed and removed from the system.
	
	\item \label{assmp:indep}
	The information particles propagate {\em independently} of each other, and their trajectories are random according to an i.i.d. random process.
\end{enumerate} 

Let $\mX$ be a finite set of constellation points on the real line: $\mathcal{X} \triangleq \{\xi_0, \xi_1, \dots, \xi_{L-1} \}$, $0 \le \xi_0 \le \dots \le \xi_{L-1} = \Delta$. 
Observing $l \in {0,1,\dots,L-1}$ with equal probability, the transmitter {\em simultaneously} releases $M$ information particles into the medium at time $X \in \mX$. The release time $X$ is assumed to be independent of the random propagation time of {\em each} of the information particles. 
Let $\{Y_m\}_{m=1}^M$ denote the $M$ arrival times of each of the information particles released at time $X$. Due to causality, we have $Y_{m} > X, m=1,2,\dots,M$. This leads to the following additive noise channel model:     
\begin{align}
	Y_{m} = {X} + Z_{m}, \quad m=1,2,\dots,M,
\label{eq:LevyChan}
\end{align}

\noindent where $Z_{m} \in \realSet$ is a random noise term representing the propagation time of the $m^{\text{th}}$ particle. Note that Assumption \ref{assmp:indep} implies that the RVs $Z_{m}$ are independent of each other. The channel model \eqref{eq:LevyChan} represents well the setting of a transmitter (e.g., a nano-scale sensor) that infrequently sends a symbol (which conveys a limited number of bits) to a receiver (e.g., a centralized molecular controller), and then remains silent for a long period. Thus, the communication has a one-shot nature.
 
To simplify the presentation, in most of this paper we restrict our attention to the case of binary modulations, i.e., $\mX = \{ 0, \Delta \}$. 
 However, the results derived in this paper can be extended to more than two elements in the set $\mX$, and to unequal a-priori probabilities. 
Let $\hat{X}$ denote the estimation of $X$ at the receiver. We denote the probability of error, when $M$ particles are used, by $P_{\varepsilon}^{(M)} \triangleq \Pr \{X \neq \hat{X} \}$.
Since all the particles are simultaneously released, and since the receiver can ignore some of the arrivals, $P_{\varepsilon}^{(M)}$ should decrease with increasing $M$, \cite{sri12, murinGlobeCom16}. 
In this paper we focus on the {\em exponential} decrease of $P_{\varepsilon}^{(M)}$ in the asymptotic limit of increasing $M$, defined by the quantity $\sf{D}$ given by:
\begin{align}
	\sf{D} \triangleq \lim_{M \to \infty} \frac{-\log P_{\varepsilon}^{(M)}}{M}.
\end{align}

\begin{remark}
	The channel \eqref{eq:LevyChan} has a single input and multiple outputs. Thus, by simultaneously releasing $M$ particles we achieve {\em receive diversity}. This motivates referring to $\sf{D}$ as the \ymc{\em system diversity gain}.	\textcolor{\changeColor}{Clearly, if $P_{\varepsilon}^{(M)}$ does not decrease {\em at least} exponentially with $M$, then the system diversity gain is $\sf{D} \mspace{-2mu} = \mspace{-2mu} 0$.}
	As the propagation of all particles is independent and identically distributed, see Assumption \ref{assmp:indep},  the channel \eqref{eq:LevyChan} can also be viewed as a single-input-multiple-output channel in which all the channel outputs experience an independent and identical propagation law.
\end{remark}

%
Note that the above description of communication over a molecular timing channel is fairly general and can be applied to different propagation mechanisms as long as Assumptions \ref{assmp:synch}--\ref{assmp:indep} are not violated. Next, we discuss the random propagation model for our channel.

\subsection{Random Propagation Model} \label{subsec:RandomPropag}

Before specifying our assumptions on the propagation model, we first define the class of weakly unimodal (quasi-concave) functions \cite[Sec. 3.4.1]{boyd-book}:

\begin{definition} \label{def:unimodal}
	A function $f(z)$ is said to be weakly unimodal if there exists a value $\zeta$ for which it is weakly monotonically increasing for $z \le \zeta$ and weakly monotonically decreasing for $z \ge \zeta$. Note that for a weakly unimodal function the maximum value can be reached for a continuous range of values of $z$.
\end{definition}

In the following we refer to functions satisfying Def. \ref{def:unimodal} as unimodal functions.
As we consider the timing channel model \eqref{eq:LevyChan}, we focus on propagation models characterized by a noise density function $f_Z(z)$ with the support $\realSet^{+}$.\footnote{Note that \cite{farsadGlobeCom16} considered a molecular timing channel with differential transmission in which the noise density is $\realSet$. Yet, for the channel model \eqref{eq:LevyChan}, the noise can have only positive values.} We further assume that the density function $f_Z(z)$ is continuous, differentiable, and unimodal. Note that we {\em do not} restrict $f_Z(z)$ to have finite first or second moments.

In many \ymc{molecular} timing channels the random propagation is characterized using densities under which the above assumptions hold. 
For instance, in diffusive molecular communications the released particles follow a random Brownian path from the transmitter to the receiver. In the case of diffusion {\em without} a drift, the RVs $Z_m$ are L\'evy-distributed \cite[Def. 1]{murinGlobeCom16}, while in the case of diffusion {\em with} a drift, the RVs $Z_m$ follow the IG distribution \cite[eq. (3)]{sri12}. 
Another example is noise with exponential density that was considered in \cite{rose:InscribedPart1}.

Next, we derive the diversity gain of the following three detectors: the ML detector, the linear (mean) detector, and the FA detector.
	
\section{\ymc{System Diversity Gain under Different Detection Methods}} \label{sec:PDG}

\subsection{The ML Detector} \label{subsec:ML}

Let $\yvec = \{ y_m\}_{m=1}^M$. The ML detector is given by the following decision rule:
\vspace{-0.1cm}
\begin{align}
	\hat{X}_{\text{ML}}(\vec{y}) = \begin{cases} 0, & \sum_{m=1}^M{\log \frac{f_Z(y_m)}{f_Z(y_m - \Delta)}}   \ge 0  \\ \Delta, &  \text{otherwise}. \end{cases} \label{eq:MLdetector}
\end{align}

\noindent For many types of $f_Z(z)$ an explicit expression for $P_{\varepsilon,\text{ML}}^{(M)}$ is not available. Yet, since the problem of recovering $x$ based on the $M$ i.i.d. realizations $\{y_m\}_{m=1}^M$ belongs to the class of binary hypothesis problems, the diversity gain is exactly the Chernoff information: 
\begin{proposition} \label{prop:ML_PDG}
	The diversity gain for the ML detector in \eqref{eq:MLdetector} is given by:
	\begin{align}
		\mspace{-8mu} \sf{D}_{\text{ML}} \mspace{-3mu} = \mspace{-1mu} - \mspace{-8mu} \min_{s: 0 \leq s \leq 1} \mspace{-3mu} \log \mspace{-2mu} \left(\int_{y=\Delta}^{\infty} \mspace{-5mu} (f_Z(y))^{s} \mspace{-3mu} \cdot \mspace{-3mu} (f_Z(y \mspace{-3mu} - \mspace{-3mu} \Delta))^{1-s} dy\right). \label{eq:ML_PDG}
	\end{align}
\end{proposition}

\begin{IEEEproof}[$\mspace{-35mu}$ Proof]
	The result follows from combining \cite[Theorem 11.9.1]{cover-book} and \cite[eq. (11.239)]{cover-book} for continuous distributions.
\end{IEEEproof}

Although the above ML detector minimizes the probability of error for equiprobable signaling, and thus it maximizes the diversity gain, it has two main drawbacks. First, it is relatively complicated to compute in low-complexity devices (e.g., nano-scale sensors). Second, the ML detector requires {\em all} the particles to arrive, which may require {\em long delays}. This is \ymc{particularly relevant} when $f_Z(z)$ has heavy tails (e.g., the L\'evy distribution). 

\subsection{The Linear Detector}

If the additive noise is Gaussian then the optimal detector is linear \cite[Ch. 3.3]{TV:05}. Even when the noise is not Gaussian, this approach can significantly improve the probability of error, as observed in \cite[Sec. IV.C.2]{sri12} for the case of additive IG noise. The main benefit of the linear detector is its simplicity; yet, it may also require long delays. Before formally defining the linear detector, we comment on the possibly destructive effect of linear detection in the case of heavy tailed $f_Z(z)$.
\begin{remark} \label{rem:destructiveLinear}
		In \cite[Thm. 1]{murinGlobeCom16} it is shown that, for the case of L\'evy-distributed propagation, a linear detector (e.g., the mean) increases the dispersion of the noise.\footnote{The dispersion of the noise is also known as its scale.} Thus, the probability of error of a linear detector is lower bounded by the probability of error of an optimal detector for the case of $M=1$, which leads to a zero diversity gain.
\end{remark}

	Consider detection based on $Y_{\text{LIN}} \mspace{-3mu} \triangleq \mspace{-3mu} \frac{1}{M} \sum_{m=1}^M \mspace{-3mu} Y_m$, and let $\Lambda_Z(\tau) \triangleq \log \E_Z \left\{ e^{\tau Z} \right\}$ denote the cumulant generating function of $Z$. Further, define $\Lambda_Z^{\ast}(v) \triangleq \sup_{\lambda} \left\{ \lambda v - \Lambda_Z(\lambda) \right\}$ to be the rate (Cram\'{e}r) function.
	The diversity gain of $\hat{X}_{\text{LIN}}(Y_{\text{LIN}})$ is stated in the following theorem.
	\begin{theorem}
		Let $f_Z(z)$ be a density with $\E \{Z\} \mspace{-3mu} = \mspace{-3mu} \mu \mspace{-3mu} < \mspace{-3mu} \infty$, and $\E \{(Z \mspace{-3mu} - \mspace{-3mu} \mu)^2 \} \mspace{-3mu} < \mspace{-3mu} \infty$. Further assume that $\Lambda_Z(\tau)$ is finite over some interval in $\realSet$. If there exists $\alpha \mspace{-3mu} \in \mspace{-3mu} (\max \{\Delta \mspace{-3mu} - \mspace{-3mu} \mu, 0\}, \Delta)$, such that $\Lambda_Z^{\ast}\left( \mu \mspace{-3mu} + \mspace{-3mu} \alpha \right) \mspace{-3mu} = \mspace{-3mu} \Lambda_Z^{\ast}\left( \mu \mspace{-3mu} - \mspace{-3mu} \Delta \mspace{-3mu} + \mspace{-3mu} \alpha \right)$, then: 
		\vspace{-0.1cm}
		\begin{align}
			\sf{D}_{\text{LIN}} = \Lambda_Z^{\ast}\left( \mu + \alpha \right). \label{eq:Lin_PDG} 
		\end{align}
		
		\vspace{-0.15cm}
		\noindent Otherwise, $\sf{D}_{\text{LIN}} = \infty$.
	\end{theorem}
	
\begin{IEEEproof}[$\mspace{-35mu}$ Proof]
	Let $P_{\varepsilon|0}^{(M)}$ and $P_{\varepsilon|\Delta}^{(M)}$ denote the probabilities of error given $x=0$ and $x=\Delta$, respectively. The diversity gain is now given by:
	\vspace{-0.1cm}
	\begin{align}
		\sf{D}_{\text{LIN}} = \min \left\{ \lim_{M \to \infty} \frac{-\log P_{\varepsilon|0}^{(M)} }{M}, \lim_{M \to \infty} \frac{-\log P_{\varepsilon|\Delta}^{(M)}}{M} \right\}. \label{eq:Lin_PDG_general}
	\end{align}
	
	\vspace{-0.15cm}
	\noindent From Cram\'{e}r's Theorem \cite[Thm. 2.2.3]{dembo-book} we have $\lim_{M\to \infty} \log \Pr \{Y_{\text{LIN}} > y_0 | X=x\} \mspace{-3mu} = \mspace{-3mu} \Lambda_Z^{\ast}\left( y_0 \right), y_0 \mspace{-3mu} > \mspace{-3mu} \mu$. 
	Similarly, Cram\'{e}r's Theorem states that $\lim_{M\to \infty} \log \Pr \{Y_{\text{LIN}} < y_\Delta | X=x\} \mspace{-3mu} = \mspace{-3mu} \Lambda_Z^{\ast}\left( y_\Delta \right), y_\Delta \mspace{-3mu} < \mspace{-3mu} \mu$. Recalling that $f_Z(z)$ is not necessarily symmetric, to maximize $\sf{D}_{\text{LIN}}$ we use the fact that the two densities differ only in a shift and require the two terms on the right-hand-side (RHS) of \eqref{eq:Lin_PDG_general} to be the same. 
	Thus, we find the point at which the right tail ($\Lambda_Z^{\ast}\left( \mu \mspace{-3mu} + \mspace{-3mu} \alpha \right)$) equals the left tail ($\Lambda_Z^{\ast}\left( \mu \mspace{-3mu} - \mspace{-3mu} \Delta \mspace{-3mu} + \mspace{-3mu} \alpha \right)$). 	
	This leads to the decision threshold $\mu + \alpha$ and to \eqref{eq:Lin_PDG}. If such a point does not exist, then the decision intervals do not overlap, which implies zero probability of error and $\sf{D}_{\text{LIN}} = \infty$.
\end{IEEEproof}
	
Next, we discuss a detector that is based on FA of a particle. \ymc{For this detector, the time gap between transmission and detection is minimal, and in for some noise distributions this detector is equivalent to the optimal ML detector.}

\subsection{The FA detector}


Let $y_{\text{FA}} \triangleq \min \left\{ y_1,y_2,\dots,y_M \right\}$. The FA detector is the ML detector based on $Y_{\text{FA}}$. Before discussing the performance of the ML detector, we note that for a fixed value of $M$, $f_{Y_{\text{FA}}|X}(y_{\text{FA}}|x)$ {\em is not necessarily unimodal}. 
The following lemma provides sufficient conditions for $f_{Y_{\text{FA}}|X}(y_{\text{FA}}|x)$ to be unimodal, for sufficiently large (yet finite) $M$.
\begin{lemma} \label{lemm:unimodalFA}
		Let $f_Z(z)$ be a unimodal density supported on $\realSet^{+}$, and $f'_Z(z)$ its derivative. 
		If there exists an $\epsilon > 0$ such that, for every $0 < z \le \epsilon$, the function $g(z) = \frac{f'_Z(z)}{f_Z^2(z)}$ is monotonically decreasing, then there exists a sufficiently large and finite $M_0$ for which the density of $Z_{\text{FA}} \mspace{-3mu} \triangleq \mspace{-3mu} \min\{Z_m\}_{m=1}^M$ is unimodal for $M>M_0$.
\end{lemma}

\begin{IEEEproof}[$\mspace{-35mu}$ Proof]
	The proof is provided in Appendix \ref{annex:proof_lemm_unimodalFA}.
\end{IEEEproof}

\begin{remark}
	Lemma \ref{lemm:unimodalFA} provides {\em sufficient} conditions for the density of the FA, $f_{Y_{\text{FA}}|X}(y_{\text{FA}}|x)$, to be unimodal when $M$ is sufficiently large. It is possible that $f_{Y_{\text{FA}}|X}(y_{\text{FA}}|x)$ will be unimodal even if these conditions do not hold. It is also possible that $f_{Y_{\text{FA}}|X}(y_{\text{FA}}|x)$ will be unimodal for values of $M$ smaller than $M_0$. In Section \ref{sec:numRes} we show that the conditions of Lemma \ref{lemm:unimodalFA} hold for the IG and for the L\'evy densities.
\end{remark}

Next, we assume that $f_{Y_{\text{FA}}|X}(y_{\text{FA}}|x)$ is unimodal for a given finite $M$, and provide the detection rule and probability of error of the FA detector. 
\begin{proposition}
	Let $f_{Y_{\text{FA}}|X}(y_{\text{FA}}|x)$ be unimodal for a given value of $M$, and let $m_Z$ denote the mode of $f_Z(z)$. \textcolor{\changeColor}{Further, let $F_Z(z)$ be the noises' CDF.} Then, the ML detector based on $y_{\text{FA}}$, is given by:
	\vspace{-0.1cm}
	\begin{align}
		\hat{X}_{\text{FA}}(y_{\text{FA}}) = \begin{cases} 0, & y_{\text{FA}} < \theta_M \\ \Delta, &  y_{\text{FA}} \ge \theta_M, \end{cases}
		\label{eq:decisionRuleFA}
	\end{align}
	
	\vspace{-0.1cm}
	\noindent where $\theta_M$, is the solution, in $\Delta \mspace{-3mu} \le \mspace{-3mu} y_{\text{FA}} \mspace{-3mu} \le \mspace{-3mu} m_Z$, of the following equation in $y_{\text{FA}}$:
	\vspace{-0.1cm}
	\begin{align}
		\frac{f_Z(y_{\text{FA}})}{f_Z(y_{\text{FA}}-\Delta)} = \left( \frac{1 - F_z(y_{\text{FA}} - \Delta)}{1 - F_z(y_{\text{FA}})} \right)^{M-1}. \label{eq:FA_decRuleEquation}
	\end{align}
	
	\vspace{-0.1cm}	
	\noindent If \eqref{eq:FA_decRuleEquation} does not have a solution, then $\theta_M = \Delta$. Furthermore, the probability of error of the FA detector is given by:
	\vspace{-0.1cm}
	\begin{align}
		\mspace{-8mu} P_{\varepsilon, \text{FA}}^{(M)} & \mspace{-3mu} = \mspace{-3mu} 0.5 \left( \left( 1 \mspace{-3mu} - \mspace{-3mu} F_z(\theta_M) \right)^M \mspace{-3mu}  + \mspace{-3mu} 1 \mspace{-3mu} - \mspace{-3mu} \left(1 \mspace{-3mu} - \mspace{-3mu} F_z(\theta_M \mspace{-3mu} - \mspace{-3mu} \Delta) \right)^M \right).
		\label{eq:errProbSymbBySymbFA}
	\end{align}

\end{proposition}

\begin{IEEEproof}[$\mspace{-35mu}$ Proof Outline]
	As $f_{Y_{\text{FA}}|X}(y_{\text{FA}}|x)$, the ML detector amounts to comparing $y_{\text{FA}}$ to a threshold, which can be found by equating the two densities. Since the two densities differ only by a shift, if \eqref{eq:FA_decRuleEquation} does not have a solution, then $\theta_M = \Delta$. Finally, \eqref{eq:FA_decRuleEquation}--\eqref{eq:errProbSymbBySymbFA} are obtained by noting that $F_{Y_{\text{FA}}|X}(y_{\text{FA}}|x) = 1 - \left(1 - F_{Z}(y-x) \right)^M$ and $f_{Y_{\text{FA}}|X}(y_{\text{FA}}|x) = M \mspace{-3mu} \cdot \mspace{-3mu} f_{Z}(y-x) \mspace{-3mu} \cdot \mspace{-3mu} \left(1 - F_{Z}(y-x) \right)^{M-1}$.	
\end{IEEEproof}

The following theorem presents the diversity gain of the FA detector.
\begin{theorem} \label{thm:FA_PDG}
	Let $f_{Y_{\text{FA}}|X}(y_{\text{FA}}|x)$ be unimodal for all $M > M_0$. Then, the diversity gain of the FA detector is given by:
	\vspace{-0.1cm}
	\begin{align}
			\sf{D}_{\text{FA}} = - \log \left(1 - F_Z(\Delta) \right). \label{eq:FA_PDG}
	\end{align}
\end{theorem}

\vspace{-0.05cm}
\begin{IEEEproof}[$\mspace{-35mu}$ Proof]
	Before proving \eqref{eq:FA_PDG} we note that if $f_{Y_{\text{FA}}|X}(y_{\text{FA}}|x)$ is unimodal for all $M > M_0$, then $\theta_M \mspace{-3mu} \to \mspace{-3mu} \Delta$ when $M \mspace{-3mu} \to \mspace{-3mu} \infty$. This follows as the RHS of \eqref{eq:FA_decRuleEquation} increases with $M$ while the left-hand-side is independent of $M$. 
	\textcolor{\changeColor}{Moreover, using the extreme value theorem \cite[Thm. 1.8.4]{extremvalue-book}, which implies that the limiting distribution of the considered densities is a Dirac delta at $x$, we conclude that the limit is indeed $\Delta$.}
	
	Next, we recall that $Z_{\text{FA}} \mspace{-3mu} = \mspace{-3mu} \min\{Z_m\}_{m=1}^M$, and let $\theta_M \mspace{-3mu} = \mspace{-3mu} \Delta \mspace{-3mu} + \delta_M, \delta_M \mspace{-3mu} \to \mspace{-3mu} 0$.
	Note that $P_{\varepsilon, \text{FA}}^{(M)}$ can be bounded as follows:
	\vspace{-0.1cm}
	\begin{align}
		\frac{1}{2}\Pr \{ Z_{\text{FA}} \ge \Delta + \delta_M \}  \le P_{\varepsilon, \text{FA}}^{(M)} \le \frac{1}{2} \Pr \{ Z_{\text{FA}} \ge \Delta \}. \label{eq:PeFA_bounds}
	\end{align}
	
	\vspace{-0.05cm}
	\noindent For the RHS of \eqref{eq:PeFA_bounds}, recalling that $\Pr \{ Z_{\text{FA}} \mspace{-3mu} \ge \mspace{-3mu} \Delta\} \mspace{-3mu} = \mspace{-3mu}  (1 \mspace{-3mu} - \mspace{-3mu} F_Z(\Delta))^M$, we write
	\begin{align}
		\lim_{M \to \infty} \frac{- \log \Pr \{ Z_{\text{FA}} \ge \Delta \}}{M}  \mspace{-3mu} = \mspace{-3mu} - \log (1 \mspace{-3mu} - \mspace{-3mu} F_Z(\Delta) ).
	\end{align}
	
	\vspace{-0.05cm}
	\noindent For the left-hand-side we have $\Pr \{ Z_{\text{FA}} \mspace{-2mu} \ge \mspace{-2mu} \Delta \mspace{-2mu} + \mspace{-2mu} \delta_M\} \mspace{-3mu} = \mspace{-3mu}  (1 \mspace{-3mu} - \mspace{-3mu} F_Z(\Delta + \delta_M))^M$. Using a Taylor expansion of $\log (1 \mspace{-3mu} - \mspace{-3mu} F_Z(\Delta + \delta_M))^M$ around $\Delta$, we obtain:
	\vspace{-0.1cm}
	\begin{align}
		& \log (1 \mspace{-3mu} - \mspace{-3mu} F_Z(\Delta + \delta_M))^M \nonumber \\
		& \qquad \mspace{-3mu} = \mspace{-3mu} M\left(\log(1 \mspace{-3mu} - \mspace{-3mu} F_Z(\Delta)) \mspace{-3mu} - \mspace{-3mu} \frac{f_Z(\Delta) \delta_M}{1 - F_Z(\Delta)} \mspace{-3mu} + \mspace{-3mu} \mathcal{O}(\delta_M^2)  \right). \nonumber 
	\end{align}
	
	\vspace{-0.05cm}
	\noindent Therefore, since $\delta_M \mspace{-3mu} \to \mspace{-3mu} 0$, we have:
	\vspace{-0.1cm}
	\begin{align}
		\lim_{M \to \infty} \frac{- \log \Pr \{ Z_{\text{FA}} \ge \Delta + \delta_M \}}{M}  \mspace{-3mu} = \mspace{-3mu} - \log (1 \mspace{-3mu} - \mspace{-3mu} F_Z(\Delta) ). \label{eq:PDG_FA_rightDist}
	\end{align}
	
	\noindent Combining \eqref{eq:PeFA_bounds}--\eqref{eq:PDG_FA_rightDist} we conclude the proof.
\end{IEEEproof}
%

\begin{remark}
	The FA, linear and ML detectors can be directly extended to the case of larger constellations, i.e., $L>2$. 
	In this case the ML detector requires comparing $L$ hypotheses. On the other hand, optimal detection based on the $Y_{\text{FA}}$ can be implemented by comparing only two hypothesizes which can be easily found based on their modes. 
	Since for the FA detector the conditional density concentrates towards $x$, for a fixed $L$ one can find large enough $M$ such that any non-zero probability of error can be achieved. This enables sending short messages of several bits using a large number of particles. 
\end{remark}

We now consider the special case in which the mode of $f_Z(z)$ is zero, e.g., the uniform or exponential densities:
\begin{theorem} \label{thm:zreomode}
	Let $f_Z(z)$ be a continuous, differentiable, and unimodal density with mode $m_Z \mspace{-3mu} \ge \mspace{-3mu} 0$ and $f_Z(z) \mspace{-3mu} = \mspace{-3mu} 0, z \mspace{-3mu} < \mspace{-3mu} m_Z$. Then, the FA and ML detectors are equivalent, namely, they have the same probability of error. 
\end{theorem}

\begin{IEEEproof}[$\mspace{-35mu}$ Proof]
	We focus on the case of $m_Z \mspace{-3mu} = \mspace{-3mu} 0$. The proof for $m_Z \mspace{-3mu} > \mspace{-3mu} 0$ follows similar lines.
	Since $m_Z = 0$, then $f_Z(y) \mspace{-3mu} \le \mspace{-3mu} f_Z(y \mspace{-3mu} - \mspace{-3mu} \Delta), y \mspace{-3mu} \ge \mspace{-3mu} \Delta$. 
	The ML detection rule can be written as:
	\vspace{-0.1cm}
	\begin{align}
	\hat{X}_{\text{ML}}(\vec{y}) = \argmax_{x} \prod_{m=1}^{M} f_Z(y_m|X=x).
	\end{align}
	
	\vspace{-0.15cm}	
	\noindent Therefore, if there exists $Y_m \mspace{-3mu} < \mspace{-3mu} \Delta$, then the ML detector declares $\hat{X}_{\text{ML}}(\vec{y}) \mspace{-4mu} = \mspace{-4mu} 0$. Otherwise it declares $\hat{X}_{\text{ML}}(\vec{y}) \mspace{-4mu} = \mspace{-4mu} \Delta$. Since testing if there exists $Y_m \mspace{-3mu} < \mspace{-3mu} \Delta$ can be implemented based on $y_{\mathrm{FA}}$, we conclude that the ML detector reduces to the FA detector in this case, so that the detectors are equivalent.
	\end{IEEEproof}
	
	\begin{corollary} \label{cor:zreomode}
		Under the conditions of Thm. \ref{thm:zreomode} $\sf{D}_{\mathrm{FA}} \mspace{-2mu} = \mspace{-2mu} \sf{D}_{\mathrm{ML}}$.
	\end{corollary}

\begin{remark}
	\ymc{While the ML detector \eqref{eq:MLdetector} is optimal for detection based on {\em all} the particle arrivals, the FA detector \eqref{eq:decisionRuleFA} is optimal for detection based {\em only on the FA} of a particle. 
	One can use order statistics theory to design optimal detectors based on the first $M_0 \mspace{-3mu} \le \mspace{-3mu} M$ particle arrivals. Yet, waiting for $M_0$ particles to arrive requires longer delays compared to the FA detection framework, and as indicated in the following section the FA detector achieves a performance very close to the performance of the ML detector.}
\end{remark}

Next, we explicitly evaluate the formulas derived above and the resulting diversity gain for several specific propagation densities: the uniform, exponential, IG, and L\'evy distributions. 

\vspace{-0.2cm}
\section{Specific Propagation Profiles \\ and Numerical Results} \label{sec:numRes}

\vspace{-0.1cm}
\subsection{The Uniform Distribution} \label{subsec:unifom}

\vspace{-0.05cm}
Let $f_Z(z) \mspace{-3mu} \sim \mspace{-3mu} \text{U}(0,b), b \mspace{-3mu} > \mspace{-3mu} \Delta$, i.e., the uniform density over $[0,b]$. 
Following Corollary \ref{cor:zreomode}, $\sf{D}_{\text{ML}} \mspace{-3mu} = \mspace{-3mu} \sf{D}_{\text{FA}} \mspace{-3mu} = \mspace{-3mu} \log \frac{b}{b-\Delta}$.
For the linear detector we do not have a closed form expression for $\Lambda_Z^{\ast}(v)$. However, we can use the moment generating function (MGF) of the uniform distribution and numerically find $\sf{D_{\text{LIN}}}$.

For instance, let $b \mspace{-3mu} = \mspace{-3mu} 1$, and consider $\Delta \mspace{-3mu} \in \mspace{-3mu} \{ 0.25, 0.5, 0.75 \}$. Table \ref{tab:uniform} details the resulting diversity gains. The table indicates large performance gains of the ML and FA detectors over linear detection.
\begin{table}[h]
		\begin{center}
		\footnotesize
		\begin{tabular}[t]{|c|c|c|c|}
			\hline
			  & $\Delta=0.25$  & $\Delta=0.5$ & $\Delta=0.75$ \\
			\hline
			\hline
			$\sf{D}_{\text{ML}}, \sf{D}_{\text{FA}}$  & 0.2879  & 0.6931  & 1.3863  \\
			\hline
			$ \sf{D}_{\text{LIN}}$  & 0.0956  & 0.4086  & 1.0798 \\
			\hline 
		\end{tabular}
		\captionsetup{font=footnotesize}
		\caption{$\sf{D}_{\text{ML}}, \sf{D}_{\text{FA}}$ and $\sf{D}_{\text{LIN}}$ for \text{U}(0,1). \label{tab:uniform}}
		\vspace{-0.65cm}
	\end{center}
\end{table}

\vspace{-0.2cm}
\subsection{The Exponential Distribution}

\vspace{-0.05cm}
Let $f_Z(z) \mspace{-3mu} \sim \mspace{-3mu} \text{Exp}(b)$, i.e., the exponential density with rate parameter $b \mspace{-3mu} > \mspace{-3mu} 0$.
An explicit evaluation of \eqref{eq:ML_PDG} results in $\sf{D}_{\text{ML}} \mspace{-3mu} = \mspace{-3mu} b\Delta$, and following Corollary \ref{cor:zreomode}, we have $\sf{D}_{\text{ML}} = \sf{D}_{\text{FA}}$. 
Considering the linear detector, for the exponential density $\Lambda_Z^{\ast}(v) \mspace{-3mu} = \mspace{-3mu} bv \mspace{-2mu} - \mspace{-2mu} 1 \mspace{-2mu} - \mspace{-2mu} \log (bv), v > 0$.
Moreover, it can be shown that the $\alpha$ in \eqref{eq:Lin_PDG} is given by:
\vspace{-0.1cm}
\begin{align}
	\alpha = \frac{1 - e^{b \Delta} (1 - b \Delta)}{(e^{b \Delta} - 1) b}.
\end{align}

\vspace{-0.15cm}
\noindent Plugging this value into \eqref{eq:Lin_PDG} results in:
\vspace{-0.1cm}
\begin{align}
	\sf{D}_{\text{LIN}} = \frac{1+e^{b \Delta}(b \Delta-1) - (e^{b \Delta}-1) \log\left(\frac{b \Delta e^{b \Delta}}{e^{b \Delta}-1}\right)}{e^{b \Delta}-1}.
\end{align} 

\vspace{-0.15cm}
As an example, let $b \mspace{-3mu} = \mspace{-3mu} 1$, and consider $\Delta \mspace{-3mu} \in \mspace{-3mu} \{ 0.5, 1.5, 2.5 \}$. Table \ref{tab:exp} details the resulting diversity gains. 
\begin{table}[h]
		\begin{center}
		\footnotesize
		\begin{tabular}[t]{|c|c|c|c|}
			\hline
			  & $\Delta=0.5$  & $\Delta=1.5$ & $\Delta=2.5$ \\
			\hline
			\hline
			$\sf{D}_{\text{ML}}, \sf{D}_{\text{FA}}$  & 0.5  & 1.5  & 2.5  \\
			\hline
			$ \sf{D}_{\text{LIN}}$  & 0.0312  & 0.2729  & 0.7216 \\
			\hline 
		\end{tabular}
		\captionsetup{font=footnotesize}
		\caption{$\sf{D}_{\text{ML}}, \sf{D}_{\text{FA}}$ and $\sf{D}_{\text{LIN}}$ for \text{Exp}(1). \label{tab:exp}}
		\vspace{-0.65cm}
	\end{center}
\end{table}

\vspace{-0.2cm}
\subsection{The Inverse-Gaussian Distribution}

Let $f_Z(z) \mspace{-3mu} \sim \mspace{-3mu} \text{IG}(\mu,b)$, i.e., the IG density with mean $\mu$ and shape parameter $b \mspace{-3mu} > \mspace{-3mu} 0$. It can be shown that for $f_Z(z) \mspace{-3mu} \sim \mspace{-3mu} \text{IG}(\mu,b)$, the conditions of Lemma \ref{lemm:unimodalFA} hold and therefore $f_Z(z)$ is unimodal for sufficiently large $M$. 
While explicitly evaluating $\sf{D}_{\text{ML}}$ seems intractable, it can be evaluated numerically. For the FA detector we use the CDF of the IG density to obtain:
\vspace{-0.05cm}
\begin{align}
	\sf{D}_{\text{FA}} \mspace{-3mu} & = \mspace{-3mu} -\log\left(1 \mspace{-3mu} - \mspace{-3mu} \Phi\left(\sqrt{\frac{b}{\Delta}} \left(\frac{\Delta}{\mu} \mspace{-3mu} - \mspace{-3mu} 1 \right) \right) \right. \nonumber \\
		& \mspace{90mu} \left. - \mspace{-1mu} e^{\frac{2b}{\mu}} \Phi \left(- \mspace{-3mu} \sqrt{\frac{b}{\Delta}} \left(\frac{\Delta}{\mu} \mspace{-3mu} + \mspace{-3mu} 1 \right) \right)  \right),
\end{align}

\vspace{-0.1cm}
\noindent where $\Phi(x)$ is the CDF of a standard Gaussian RV. Finally, for the IG density $\Lambda_Z(\tau) \mspace{-3mu} = \mspace{-3mu} \frac{b}{\mu}\left(1 \mspace{-3mu}- \mspace{-3mu}\sqrt{1 \mspace{-3mu} - \mspace{-3mu} \frac{2\mu^2 \tau}{b}} \right), \tau \mspace{-3mu} \le \mspace{-3mu} \frac{b}{2\mu^2}$. Explicitly calculating $\Lambda_Z^{\ast}(v)$ for the IG density we obtain:
\begin{align}
	\Lambda_Z^{\ast}(v) \mspace{-3mu} = \mspace{-3mu} \frac{b \left(-\mu^2 + 2 \mu \left(\frac{\mu}{v} -1 \right)v + v^2 \right)}{2 \mu^2 v}. \label{eq:IG_RateFunc}
\end{align}

\noindent While for the IG density finding an explicit expression for $\alpha$ seems intractable, it can be found numerically using \eqref{eq:IG_RateFunc}. Table \ref{tab:IG} details the diversity gain for $\mu \mspace{-3mu} = \mspace{-3mu} 1, b \mspace{-3mu} = \mspace{-3mu} 1$, and $\Delta \mspace{-3mu} \in \mspace{-3mu} \{0.5, 1, 1.5 \}$. 
\begin{table}[h]
		\begin{center}
		\footnotesize
		\begin{tabular}[t]{|c|c|c|c|}
			\hline
			  & $\Delta=0.5$  & $\Delta=1$ & $\Delta=1.5$ \\
			\hline
			\hline
			$\sf{D}_{\text{ML}}$  & 0.4766  & 1.1070  &  1.6657 \\
			\hline
			$\sf{D}_{\text{FA}}$  & 0.4541  & 1.1029  & 1.6648  \\
			\hline
			$ \sf{D}_{\text{LIN}}$  & 0.0308  & 0.1180  & 0.2499  \\
			\hline 
		\end{tabular}
		\captionsetup{font=footnotesize}
		\caption{$\sf{D}_{\text{ML}}, \sf{D}_{\text{FA}}$ and $\sf{D}_{\text{LIN}}$ for \text{IG}(1,1). \label{tab:IG}}
		\vspace{-0.55cm}
	\end{center}
\end{table}


\vspace{-0.05cm}
\subsection{The L\'evy Distribution}

We last consider the L\'evy density, $f_Z(z) \mspace{-3mu} \sim \mspace{-3mu} \text{L}(\mu,b)$, with a {\em location} parameter\footnote{Note that $\mu$ is {\em not} the mean, as the mean of the L\'evy density is $\infty$.} $\mu$ and a scale parameter $b \mspace{-3mu} > \mspace{-3mu} 0$.
Similarly to the IG density, it can be shown that for $f_Z(z) \mspace{-3mu} \sim \mspace{-3mu} \text{L}(\mu,b)$, the conditions of Lemma \ref{lemm:unimodalFA} hold and therefore $f_Z(z)$ is unimodal for sufficiently large $M$. Moreover, explicitly evaluating $\sf{D}_{\text{ML}}$ seems intractable, yet, it can be evaluated numerically. For the FA detector we use the CDF of the L\'evy density to obtain $\sf{D}_{\text{FA}} \mspace{-3mu} = \mspace{-3mu} - \log \left(1 \mspace{-3mu} - \mspace{-3mu} \text{erfc}\left(\sqrt{\frac{b}{2\Delta}} \right) \right)$, where $\text{erfc}(\cdot)$ is the complementary error function. As for the linear detector we recall Remark \ref{rem:destructiveLinear} which states that for L\'evy-based propagation $\sf{D}_{\text{LIN}} \mspace{-3mu} = \mspace{-3mu} 0$.

Table \ref{tab:Levy} details the system diversity gain for $\mu \mspace{-3mu} = \mspace{-3mu} 0,b \mspace{-3mu} = \mspace{-3mu} 1$, and $\Delta \mspace{-3mu} \in \mspace{-3mu} \{0.5, 1, 1.5 \}$.
\begin{table}[h]
		\begin{center}
		\footnotesize
		\begin{tabular}[t]{|c|c|c|c|}
			\hline
			  & $\Delta=0.5$  & $\Delta=1$ & $\Delta=1.5$ \\
			\hline
			\hline
			$\sf{D}_{\text{ML}}$  &  0.1791 &  0.3828 &  0.5350 \\
			\hline
			$\sf{D}_{\text{FA}}$  & 0.1711  & 0.3817  &  0.5348  \\
			\hline 
		\end{tabular}
		\captionsetup{font=footnotesize}
		\caption{$\sf{D}_{\text{ML}}$ and $\sf{D}_{\text{FA}}$ for \text{L}(0,1). \label{tab:Levy}}
		\vspace{-0.55cm}
	\end{center}
\end{table}

It can be clearly observed that the performance gap between the FA and ML detectors is very small for both the IG and L\'evy densities. 

\vspace{-0.1cm}
\section{Conclusion} \label{sec:conc}

We have studied one-shot communication over \ymc{molecular} timing channels assuming that $M$ information particles are simultaneously released and that their propagation follows a unimodal density. 
We defined the system diversity gain $\sf{D}$ to be the exponential rate of decrease of $P_{\varepsilon}^{(M)}$ when $M$ grows asymptotically large. We then derived closed form expressions for the $\sf{D}$ achievable by three detectors: the optimal ML detector, a linear detector, and the FA detector. 
\ymc{We showed that the FA detector achieves a diversity gain very close to that of the ML detector while being simpler and having significantly shorter delays.
In particular, for delay densities where the mode of the density is zero, the FA detector is optimal. Even for other delay densities, such as the IG or the L\'evy, our numerical evaluations show that the FA detector achieves performance very close to that of the ML detector, and significantly outperforms the linear detector.} Thus, the FA detector constitutes a low-complexity near-ML detection framework for one-shot communication over timing channels. 

\appendices

\vspace{-0.1cm}
\section{Proof of Lemma \ref{lemm:unimodalFA}} \label{annex:proof_lemm_unimodalFA}

A unimodal density supported on $\realSet^{+}$ belongs to one of the following classes of densities: 1) Unimodal densities with mode $m_Z \mspace{-3mu} = \mspace{-3mu} 0$. 2) Unimodal densities with $m_Z \mspace{-3mu} > \mspace{-3mu} 0$, and $\lim_{z \to 0^+} f_Z(z) \mspace{-3mu} = \mspace{-3mu} \tau \mspace{-3mu} > \mspace{-3mu} 0$. 3) Unimodal densities with $m_Z \mspace{-3mu} > \mspace{-3mu} 0$, and $\lim_{z \to 0^+} f_Z(z) \mspace{-3mu} = \mspace{-3mu} 0$.
		%
		%
	%
We next show that densities from the first two classes are unimodal for sufficiently large $M$ regardless of the conditions of Lemma \ref{lemm:unimodalFA}. Then, we show that the conditions stated in Lemma \ref{lemm:unimodalFA} ensure that densities from the third class are unimodal for sufficiently large $M$. 

From Def. \ref{def:unimodal}, if a unimodal density has more than a single maximum, then the maximum must be a continuous interval. Thus, the derivative of the density changes its sign at most once. Next, recall that $f_{Z_{\mathrm{FA}}}(z) \mspace{-3mu} = \mspace{-3mu} M \mspace{-3mu} \cdot \mspace{-3mu} f_Z(z) \mspace{-3mu} \cdot \mspace{-3mu} (1 \mspace{-3mu} - \mspace{-3mu} F_Z(z))^{M-1}$. Hence, the derivative of $f_{Z_{\mathrm{FA}}}(z)$ is given by:
\vspace{-0.15cm}
\begin{align}
	f^{'}_{Z_{\mathrm{FA}}}(z) \mspace{-3mu} & = \mspace{-3mu} M \left( f^{'}_Z(z)(1 \mspace{-3mu} - \mspace{-3mu} F_Z(z))^{M-1} \mspace{-3mu} \right. \nonumber \\
	& \mspace{70mu} \left. - f^2_Z(z)(M-1)(1 \mspace{-3mu} - \mspace{-3mu} F_Z(z))^{M-2} \right).
\end{align}

\vspace{-0.15cm}
Setting $f^{'}_{Z_{\mathrm{FA}}}(z) \mspace{-3mu} = \mspace{-3mu} 0$ we obtain conditions indicating when $f_{Z_{\mathrm{FA}}}(z)$ decreases:
\vspace{-0.15cm}
\begin{align}
	f^{'}_{Z_{\mathrm{FA}}}(z) \mspace{-3mu} \le \mspace{-3mu} 0 \quad \Leftrightarrow \quad \frac{f^{'}_Z(z)(1 \mspace{-3mu} - \mspace{-3mu} F_Z(z))}{f^2_Z(z)} \mspace{-3mu} \le \mspace{-3mu} M-1. \label{eq:deriv_negative}
\end{align} 

\vspace{-0.15cm}
\noindent For densities that belong to the first class we have $f^{'}_Z(z) \mspace{-3mu} \le \mspace{-3mu} 0$. Therefore, as $(1 \mspace{-3mu} - \mspace{-3mu} F_Z(z))$ and $f^2_Z(z)$ are positive, $f^{'}_{Z_{\mathrm{FA}}}(z)$ is non-increasing and unimodal for {\em any} $M$. 

For the second class we note that in the range $0 \mspace{-3mu} < \mspace{-3mu} z \mspace{-3mu} \le \mspace{-3mu} m_Z$, $f^2_Z(z) \mspace{-3mu} \ge \mspace{-3mu} \tau^2$. Thus, $\frac{f^{'}_Z(z)(1 \mspace{-3mu} - \mspace{-3mu} F_Z(z))}{f^2_Z(z)}$ is positive and bounded, and by choosing $M$ large enough $f_{Z_{\mathrm{FA}}}(z)$ is decreasing for any $z$ and therefore unimodal. 

Finally, for densities in the third class, $\lim_{z \to 0^+} f_Z(z) \mspace{-3mu} = \mspace{-3mu} 0$, and since $f_Z(z)$ is assumed to be differentiable we obtain:\footnote{Recall that since $m_Z \mspace{-3mu} > \mspace{-3mu} 0$, then there exists an $\epsilon>0$ such that $0 \mspace{-3mu} \le \mspace{-3mu}f^{'}_Z(z), 0 \mspace{-3mu} < \mspace{-3mu} z \mspace{-3mu} < \mspace{-3mu} \epsilon$.}
\vspace{-0.1cm}
\begin{align}
	\lim_{z \to 0^+} \frac{f^{'}_Z(z)(1 \mspace{-3mu} - \mspace{-3mu} F_Z(z))}{f^2_Z(z)} \mspace{-3mu} = \mspace{-3mu} \infty.
\end{align}

\vspace{-0.15cm}
\noindent Since $(1 \mspace{-3mu} - \mspace{-3mu} F_Z(z))$ is monotonically decreasing with $z$, requiring that $\frac{f^{'}_Z(z)}{f^2_Z(z)}$ will decrease monotonically for $0 \mspace{-3mu} < \mspace{-3mu} z \mspace{-3mu} < \mspace{-3mu} \epsilon$ ensures that $\frac{f^{'}_Z(z)(1 \mspace{-3mu} - \mspace{-3mu} F_Z(z))}{f^2_Z(z)}$ will also be monotonically decreasing. In such case there is a $z_0$ for which:
\vspace{-0.1cm}
\begin{align}
	\frac{f^{'}_Z(z)(1 \mspace{-3mu} - \mspace{-3mu} F_Z(z))}{f^2_Z(z)} \begin{matrix} z \mspace{-3mu} < \mspace{-3mu} z_0 \\ \gtrless \\ z \mspace{-3mu} > \mspace{-3mu} z_0 \end{matrix} M-1.
\end{align}

\vspace{-0.1cm}
Hence, the density is unimodal for all $M \mspace{-3mu} > \mspace{-3mu} M_0$, where $M_0$ is given by $M_0 \mspace{-4mu} = \mspace{-4mu} \left\lceil \mspace{-3mu} \frac{f^{'}_Z(\xi)(1 \mspace{-3mu} - \mspace{-3mu} F_Z(\xi))}{f^2_Z(\xi)} \right\rceil \mspace{-4mu} + \mspace{-4mu} 1$
%
\noindent and $\xi \mspace{-4mu} = \mspace{-3mu} \argmax_z \frac{f^{'}_Z(z)(1 \mspace{-3mu} - \mspace{-3mu} F_Z(z))}{f^2_Z(z)}, z \mspace{-3mu} > \mspace{-3mu} \epsilon$.
 

\vspace{-0.05cm}
\ymc{\section*{Acknowledgment}

\vspace{-0.05cm}
The authors would like to thank Andrea Montanari for comments which greatly simplified the proof of Thm. \ref{thm:FA_PDG}.}

\vspace{-0.1cm}
\bibliographystyle{IEEEtran}
\bibliography{IEEEabrv,MCRxDesign}

\end{document}